\documentstyle[11pt]{article}
\input epsf
\begin{document}
\baselineskip 20pt
\thispagestyle{empty}
\rightline{CU-TP-690}
\vskip 2.5cm
\centerline{\large\bf QUANTUM STABILITY OF ACCELERATED BLACK
HOLES\footnote{This work is supported in part by U.S. Department of Energy.}}
\vskip 1.5cm
\centerline{{ Piljin Yi}\footnote{e-mail address:
piljin@cuphyc.phys.columbia.edu}}
\vskip 5mm
\centerline{Physics Department, Columbia University}
\centerline{New York, New York, 10027, U.S.A}
\vskip 2.5cm
\centerline{\bf ABSTRACT}
\begin{quote}
We study quantum aspects of the accelerated black holes in some detail.
Explicitly shown is the fact that a uniform acceleration stabilizes certain
charged black holes against the well-known thermal evaporation. Furthermore,
a close inspection of the geometry reveals that this is possible only for
near-extremal black holes and that most nonextremal varieties continue to
evaporate with a modified spectrum under the acceleration. We also introduce
a two-dimensional toy model where the energy-momentum flow is easily obtained
for general accelerations, and find the behavior to be in accordance with the
four-dimensional results. After a brief comparison to the classical system of
a uniformly accelerated charge, we close by pointing out the importance of
this result in the WKB expansion of the black hole pair-creation rate.
\end{quote}
\vskip 1.5cm
\leftline{PACS\#: 04.70.Dy, 03.70.+k, 04.20.Gz}

\newpage

\section{Introduction}

The Hawking radiation from the stationary black hole \cite{hawking}, is
by now a well-understood phenomenon within the semiclassical framework. Many
quantum mechanical concepts, such as energy quanta and the occupation numbers,
turned out to be coordinate-dependent ones, and this ambiguity leads to the
particle creation in the presence of the event horizon. A black
hole that has nonzero $T_{BH}$, emits thermal radiations and thereby loses its
mass steadily. That is, unless  it is eventually stabilized by a
conserved local charge inside. The canonical example of the latter is
given by the well-known Reissner-Nordstrom (RN) black holes \cite{gravity},
the minimal variant of which, namely the extremal case, has vanishing $T_{BH}$.

It turns out that there is another type of situation \cite{none} when the
Hawking's thermal radiation vanishes, and that despite nonzero $T_{BH}$. To
understand this, we need to recall a related phenomenon of the so-called
acceleration heat bath \cite{bath}\cite{unruh}. Through a similar quantum
effect as in black hole radiance, the usual Minkowskian vacuum feels like a
heat bath to uniformly accelerated Rindler observers, with the acceleration
temperature $T_{A}$ equal to the acceleration multiplied by $\hbar/2\pi$.

Now suppose a RN black hole of nonzero Hawking temperature $T_{BH}$ is
undergoing a uniform acceleration.
Furthermore, suppose that $T_{BH}$ equals $T_{A}$. Then,
co-moving Rindler observers  feels not only the
thermal radiation from the black hole but also the acceleration heat bath of
the same temperature. In effect, it is as if one put a small
blackbody of temperature $T_{BH}$ inside a large thermal cavity of
temperature $T_A=T_{BH}$. There will not be any net flow of energy, since
everything is in thermal equilibrium, which means that the co-moving
Rindler observers find the black hole stable against Hawking's thermal
evaporation.

Of course, these one-loop effects are notoriously observer-dependent, and
must be analyzed carefully within the semiclassical framework. In particular,
the acceleration heat bath is not a real entity to inertial observers (for the
good reason that the acceleration heat bath is a particular manifestation
of the usual Minkowski vacuum,
 as perceived by the Rindler observers only), and
thus it is rather difficult to imagine why the black hole should not evaporate
as seen by asymptotic inertial observers. But the point is that the Hawking
radiation and the acceleration heat bath are of the same theoretical origin,
and in a sense equally observer-dependent. The difference is only in the
choice of the initial conditions. In Ref.~\cite{none}, the author showed
that in fact the asymptotic inertial observers find neither
the acceleration heat bath (which was
expected) nor the Hawking's thermal radiation (which was not): The black
hole does not evaporate at all.

\vskip 5mm
In this article, we wish to investigate the semiclassical physics of
accelerated RN black holes in more detail, and consider the consequences.
After presenting the corresponding Ernst geometry  in Section 2, we begin
with a toy model in Section 3. The toy model has all the relevant features
of the Ernst spacetime, yet allows us to find the covariant one-loop
energy-momentum expectation value in a straightforward manner. In section
4, we return to the four dimensional Ernst spacetime and rederive the
trivial late-time Bogolubov transformation. In both cases, the conclusion
is that the black hole in question is stabilized despite the nonzero $T_{BH}$.
In Section 5, the matter of the evolution is addressed by asking when
$T_A$ can be larger than $T_{BH}$. We find the most nonextremal black hole
continues radiates while being accelerated and that the semiclassical
evolution always stops with near-extremal black holes.
In Section 6, after a comparison to the superficially similar
(classical) system of a uniformly accelerated charged particle
\cite{boulware}, we discuss the implication of our findings
in the context of RN black hole pair-creation.

\section{Uniformly Accelerated RN Black Holes}

The geometry we will consider is that of the Ernst metric \cite{ernst}. This
represents two oppositely charged magnetic RN black holes uniformly
accelerated away from each other, where the driving force is an external
magnetic field that diminishes rapidly away from the axis of symmetry.
Let us first write down the Ernst metric in a new coordinate system.
\begin{equation}
g=\frac{\Lambda^2}{(1+rAx)^2}\biggl\{-F(r)\,ds^2+F(r)^{-1}dr^2+r^2\,G(x)^{-1}
dx^2+r^2\,G(x)\,\Lambda^{-4}d\phi^2\biggr\}. \label{geometry}
\end{equation}
Starting with the Ernst metric in Ref.~\cite{gauntlett}, we performed a
coordinate redefinition $r=-1/Ay$ and also rescaled the time coordinate by
$A$. The new coordinate $r$ plays the role of the usual radial coordinate
only near the black hole horizon, as is easily seen from the form of $F(r)$.
\begin{equation}
F(r)\equiv -A^2r^2\,G(-1/Ar)=(1-\frac{r_-}{r})(1-\frac{r_+}{r}-A^2r^2).
\end{equation}
$\Lambda$ is a function of $r$ and $x$ in general:
\begin{equation}
\Lambda\equiv \biggl\{1+\frac{Bx}{2}\sqrt{r_+ r_-}\,\bigg\}^2+\frac{
B^2r^2}{4(1+rAx)^2}\,G(x),
\end{equation}
where $B$ is approximately the magnetic field strength that
drives the uniform acceleration.

It takes some investigation to discover that the Killing coordinate $s$
here is actually a Rindler time coordinate \cite{gauntlett}, so that the
``static''  observers are in fact
uniformly accelerated and thus confined within the
Rindler wedges {\large\it LR} and {\large\it RR} in figure 1: They are
Rindler observers.
\vskip 1cm
\begin{center}
\leavevmode
\epsfysize 3in \epsfbox{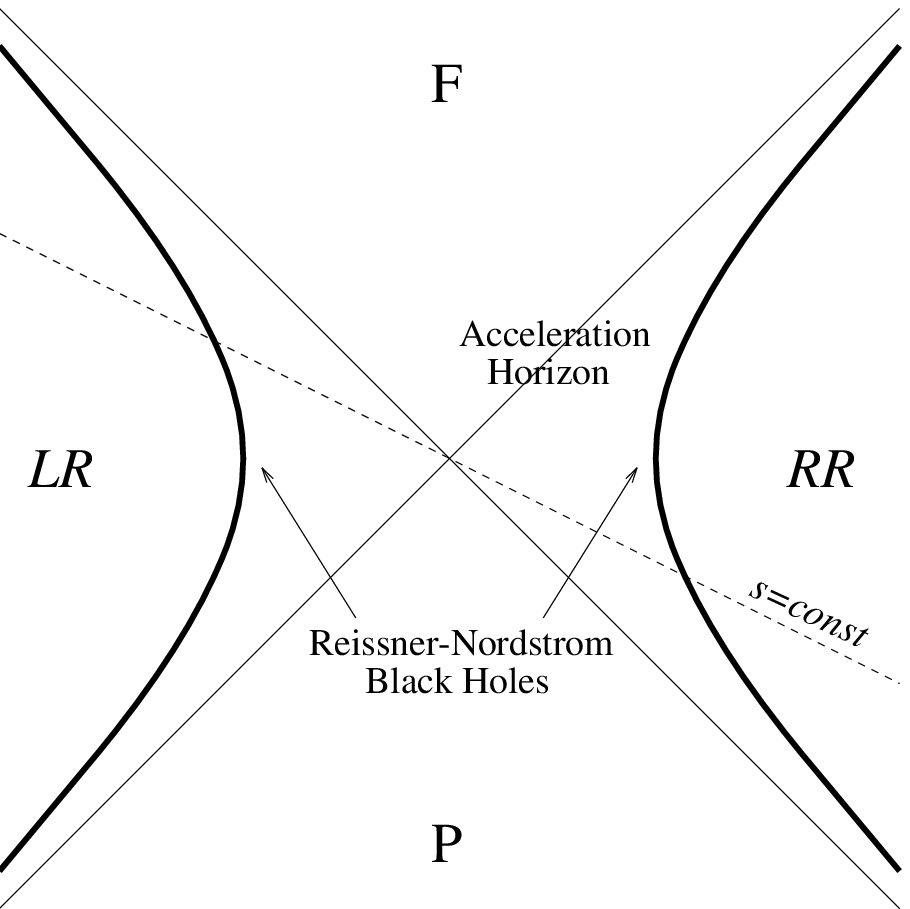}
\end{center}
\begin{quote}
{\bf Figure 1:} {\small A schematic diagram for a pair of uniformly
accelerated charged black holes. The black holes are represented by two
hyperbolic world lines in each Rindler wedges.}
\end{quote}

\vskip 1cm
The same quartic polynomial $G$ appears in all components of the metric.
Call the four roots of it, $\xi_1$, $\xi_2$, $\xi_3$, $\xi_4$
in the ascending order. Then, $r=\tilde{r}_+ \equiv -1/\xi_2A$ and $r=r_{A}
\equiv-1/\xi_3A$ are respectively the outer or event horizon of the black
holes and the acceleration horizon. On the transverse part of the geometry
with the coordinates $x$ and $\phi$, there could be conical
singularities at $x=\xi_3$ and $x=\xi_4$, which are easily resolved
by adjusting the period of $\phi$ and also putting a constraint between
parameters of the metric:
\begin{equation}
{G'(x)}/{\Lambda^2}\biggr\vert_{x=\xi_3}=
{G'(x)}/{\Lambda^2}\biggr\vert_{x=\xi_4}.
\end{equation}
When the black holes are of sufficiently small size ($r_\pm A \ll 1$),
this constraint is easily seen to reproduce the Newton's second law
\cite{gauntlett}.

\vskip 5mm
Although the geometry is rather complicated, one can easily recover the
familiar structures if the relative size of the black holes is small
($r_\pm A\ll 1$). For instance, as the acceleration decreases ($A\ll
1/r_+$), the metric resembles that of a static RN black hole in an
increasingly larger region of the spacetime. For this limit, simply ignore
terms proportional to the acceleration or the external magnetic field, and
introduce an angular coordinate $\theta=\cos^{-1} x$:
\begin{equation}
g\simeq -\tilde{F}(r)\,ds^2+\tilde{F}(r)^{-1}dr^2+r^2d\theta^2+r^2\sin^2
\theta\,d\phi^2, \qquad \tilde{F}(r)\equiv
(1-\frac{r_+}{r})(1-\frac{r_-}{r}), \label{bh}
\end{equation}
In this limit, therefore, $(r_++r_-)/2$ is the mass of the RN black hole while
$\sqrt{r_+r_-}$ is the magnetic charge inside.
On the other hand, far away from the black hole ($rA\sim 1$), one may
introduce a new set of spatial coordinates $\eta$ and $\rho$ by
\begin{equation}
A^2\eta^2\simeq \frac{1-r^2A^2}{(1+rAx)^2},\qquad
A^2\rho^2\simeq A^2r^2\frac{1-x^2}{(1+rAx)^2}.
\end{equation}
In these coordinates, the geometry far away from the black hole can be
represented by the following approximate metric,
\begin{equation}
g\simeq \tilde{\Lambda}^2\,(-A^2\eta^2ds^2+d\eta^2+d\rho^2)+
\tilde{\Lambda}^{-2}\rho^2
d\phi^2,\qquad \tilde{\Lambda}\equiv 1+\frac{B^2\rho^2}{4}, \label{melvin}
\end{equation}
which is simply the magnetic Melvin universe written in a Rindler-type
coordinate system. From the curvature of each approximate metric, it is
easy to see that the transition from the black hole geometry (\ref{bh}) to the
Melvin geometry (\ref{melvin}) must occurs at $r^3\sim r_+/B^2$ and $\eta
\simeq 1/A$, $\rho \ll 1/A$.

We expect $\eta^{-1}$ to play the role of the absolute acceleration of the
local Rindler observers,
and this tells us that the small black hole and co-moving observers nearby
must be experiencing an acceleration $\simeq A$. For freely falling inertial
observers far away from the black holes, on the other hand, the Rindler
coordinate above is not appropriate, and we need to introduce yet another
set of coordinates as follows,
\begin{equation}
T\equiv \eta \sinh As,\qquad Z\equiv \eta \cosh As, \label{mink}
\end{equation}
which are precisely the analogue of the ordinary Minkowski coordinate.

\vskip 5mm
We finally come to the most essential part, namely the causal structure.
Unlike  a stationary black hole, this geometry possesses
an extra horizon outside the event horizon. This in effect divides
asymptotic infinities into two different classes: There are asymptotic
infinities corresponding to $x=-1/Ar=\xi_3$, and those corresponding to
$x=-1/Ar>\xi_3$. To reach the first, one need not cross the acceleration
horizon at $r=-1/\xi_3A$, hence, they belong to the Rindler wedges
{\large\it LR} or {\large\it RR}. The second class on the other hand should
belong to  {\large F} or {\large P}.

Now the point is that as long as we are concerned with a radiative process,
we may as well safely ignore the first class of the infinities.
The reason is simple: Most spacetime trajectories of quanta,
being either time-like or null, will eventually cross acceleration horizon
into the region {\large F} which is inaccessible to the Rindler observers
\cite{boulware}. In order to remain within the Rindler wedge forever, the
particle must either itself have magnetic charge\footnote{Throughout the
article, we shall assume that there exits no light magnetic particle.}
or be directed exactly parallel to the axis of the uniform acceleration.
It is then easy to draw the Penrose diagram of this truncated spacetime.
See figure 2.
\vskip 1.5cm
\begin{center}
\leavevmode
\epsfysize 4in \epsfbox{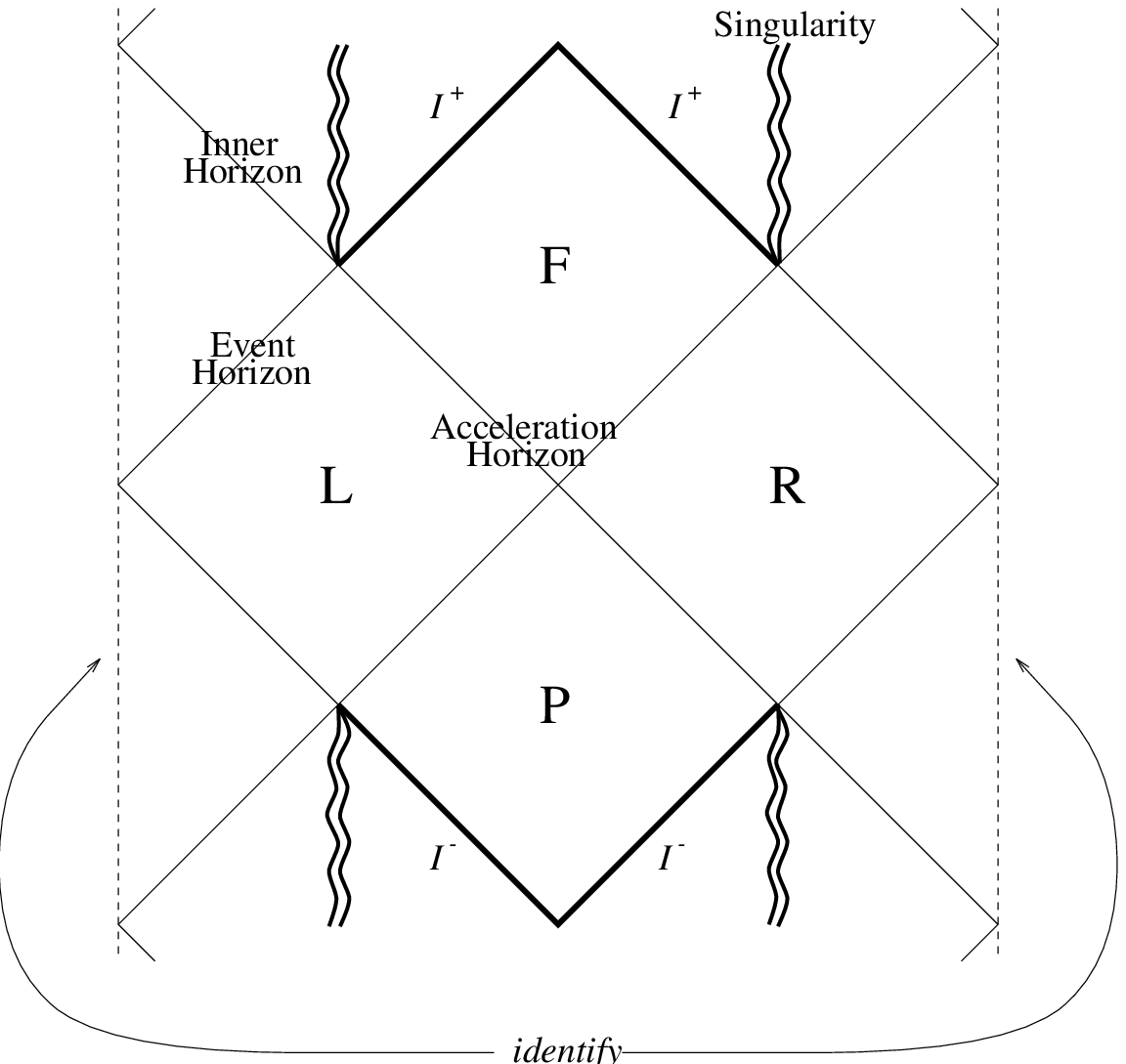}
\end{center}
\begin{quote}
{\bf Figure 2:} {\small Penrose diagram of the Ernst spacetime
with the Rindler infinities at $x=\xi_3=-1/Ar$ excised. The other
asymptotic infinities are indicated by the bold straight lines.}
\end{quote}

\vskip 5mm\noindent
Note that this Ernst spacetime actually represents a wormhole with the two
mouths (or black holes) identified at the bifurcation surface. This geometry
would be highly relevant if the black hole pairs are created via quantum
tunneling mentioned above. However, when one considers black holes made
by gravitational collapse, only the future horizons would be present.
Neither the past horizons nor the bifurcation surface are physical.
Naturally then, one should raise the question in what sense quantizing
a field in this Ernst geometry shed lights on the case of a physical black
hole under uniform acceleration.

In this paper, we shall rely on the following point in order to argue
that the study of the Hawking effect on this geometry is sufficient to
understand what really happens in more physical circumstances. First
of all, in order to establish the absence of the steady Hawking radiation,
it is sufficient to find the trivial late-time Bogolubov transformation .
Furthermore, the {\it late-time} Bogolubov transformation that normally
leads to the steady flux of thermal radiation appears by itself
quite insensitive to the history, as the reader may recall from the
example of the Hawking radiation from a freely falling black hole. To
incorporate the actual history such as the fact the black hole is made by
 gravitational collapse, one may simply puts different vacuum states for
ingoing and outgoing modes. And this only {\it after} deriving the leading
late-time Bogolubov transformation in the eternal black hole geometry which
is a maximally extended version complete with the bifurcation surface as well
as the unphysical region containing other asymptotic universes.
An independent argument that strongly supports this point of view is
presented at the end of  section 4.

But before delving into this four-dimensional problem, it is quite
instructive to consider a two-dimensional toy model where a more explicit
calculation should be possible for conformally coupled quantum fields.
Here one can start from the Unruh type physical initial condition and derive
the one-loop radiation directly from the path integral approach, instead of
deriving the Bogolubov transformation first. As we shall see below, the
derivation involves neither the bifurcation surface nor the past
event horiozns, yet the result is again consistent with the four-dimensional
one for  the eternal Ernst geometry. This clearly
reinforces our expectation that the trivial late-time Bogolubov
transformation in the maximally extended geometry leads to vanishing
Hawking radiation.

\section{Toy Model}

The Ernst geometry introduced above is only axially symmetric, which makes
a separation of variable impossible even for the simplest field equation.
In Ref.~\cite{none}, this difficulty is side-stepped by setting up the
problem in such a way that only the approximate behaviors of the eigenmodes
near each horizon matter. Although, this approach turned out to be very
powerful, it is not suitable for more general situations where the
acceleration heat bath may not be in equilibrium with
the black hole radiance.

On the other hand, if we only want to understand the system qualitatively,
there is an easy way out: consider a vastly simplified model that has rough
characters of such accelerated RN black holes, namely a RN black hole
in De Sitter universe.\footnote{The Hawking effect in the De Sitter universe
was first studied in Ref.~\cite{gibbons}.} Written in the static
coordinate, the metric is given by,
\begin{equation}
\tilde{g}=-{\cal F}(R)\,dt^2+{\cal F}(R)^{-1}dR^2+R^2d\Omega^2,\qquad
{\cal F}(R)=1-\frac{2M}{R}+\frac{Q^2}{R^2}-H^2R^2
\end{equation}
where $Q$ is the electromagnetic charge inside the black hole. For
sufficiently small cosmological constant $3H^2$, the geometry at small $R$
is just that of the ordinary RN black hole.

This geometry has three horizons, outermost of which, the so-called De Sitter
horizon, can be thought of as the analogue of the acceleration horizon of the
Ernst geometry. Relative coordinate transformations between inertial observers
near respective horizons are similar to those of the Ernst geometry, provided
that the surface gravities in one case are identified with those of the other.
As was asserted in Ref.~\cite{none}, it is precisely the Bogolubov
transformation between such two sets of inertial observers in the Ernst
geometry that are relevant for the asymptotic Hawking radiation. By considering
this toy model, therefore, we should be able to capture a qualitative
semiclassical character of the accelerated black holes.
\vskip 1cm
\begin{center}
\leavevmode
\epsfysize=2in \epsfbox{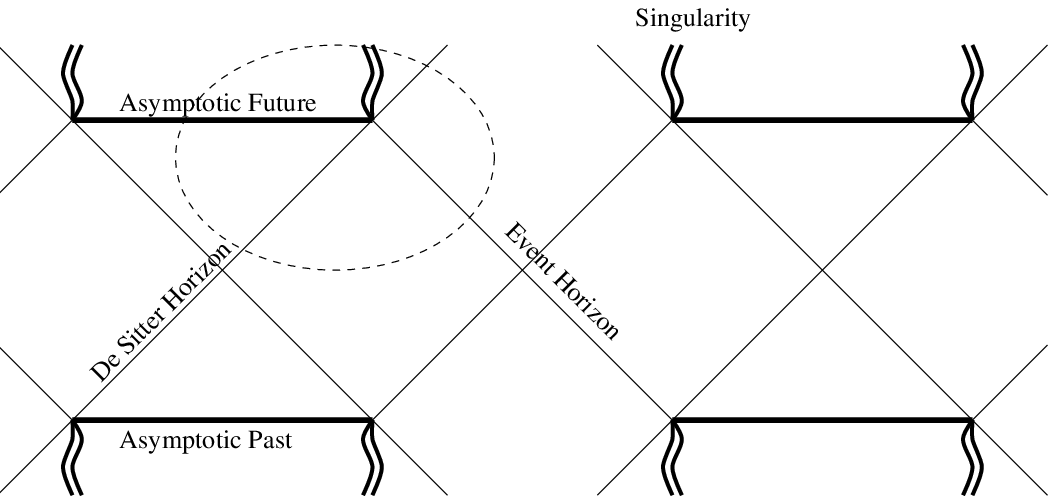}
\end{center}
\begin{quote}
\vskip 5mm
{\bf Figure 3:} {\small Penrose diagram of the charged black hole in De
Sitter universe. There is no distinction between the null and the timelike
infinities, for the ambient universe is inflating. The relevant region
inside the broken curve is redrawn in figure 4 below.}
\end{quote}

\vskip 5mm
The physical reason why we could expect a similar behavior from this
model as in uniformly accelerated black holes, is quite simple:
the inflation of the ambient De Sitter universe. Although we cannot
say that the black hole undergoes an acceleration any more, the asymptotic
observer perceives the black hole as being uniformly ``accelerated'' away
because the ambient space itself is expanding exponentially.
Semiclassically, the acceleration heat bath of the Ernst geometry would
be replaced by the thermal behavior associated with the De Sitter horizon,
but the net effect is again that certain non-inertial observers near the
black hole find a thermal equilibrium whenever the respective surface
gravities $\kappa_{bh}$ and $\kappa_{dS}$ coincides,
\begin{equation}
\kappa_{bh}=\frac{{\cal F}'(R_{bh})}{2},\qquad
\kappa_{dS}=-\frac{{\cal F}'(R_{dS})}{2},
\end{equation}
where $R_{bh}$ and $R_{dS}$ are the radii of the black hole event horizon
and the De Sitter horizon respectively.
As is easily seen from the form of $\cal F$, $\kappa_{dS}$ reduces to the
Hubble constant $H$ for the sufficiently small black hole ($M$, $Q
\rightarrow 0$).

\vskip 5mm
The spherical symmetry gives us another opportunity to simplify the problem:
the dimensional reduction. In effect, we shall consider the problem in the
S-wave sector. Quantizing an effectively two-dimensional conformal field
$\phi=\phi(t,R)$ that is coupled to the reduced two-dimensional
metric $\tilde{g}^{(2)}$:
\begin{equation}
\int {\cal L}=\int \sqrt{-\tilde{g}^{(2)}}\;(\nabla\phi)^2,\qquad
\tilde{g}^{(2)}=-{\cal F}(R)\,dt^2+{\cal F}(R)^{-1}dR^2,
\end{equation}
various quantities of interest may be obtained from the conformal anomaly
alone \cite{cghs}, as we will do in this section.

Introducing a tortoise coordinate $r_*$, between the event horizon and the
De Sitter horizon,
\begin{equation}
{r_*}=\int^R d\tilde{R}\,\frac{1}{{\cal F}(\tilde{R})},
\end{equation}
and defining the retarded and the advanced null coordinates $\tilde{u}=t-r_*$
and $\tilde{v}=t+r_*$, the energy-momentum conservation translates to
certain inhomogeneous differential equations for $T_{\tilde{u}\tilde{u}}$
and $T_{\tilde{v} \tilde{v}}$, with the inhomogeneous term proportional
to the conformal anomaly $\sim \langle T_{\tilde{u}\tilde{v}}\rangle$ or
equivalently the two-curvature $\sim {\cal F}''$. Then, it is an elementary
exercise to solve them for the outward energy flux  \cite{jaemo}:
\begin{equation}
\langle T_{\tilde{u}\tilde{u}}\rangle \sim \hbar\,
[\partial_{\tilde{u}}^2\rho-\partial_{\tilde{u}}\rho\partial_{\tilde{u}}
\rho]+t_{\tilde{u}\tilde{u}}(\tilde{u}), \qquad \rho\equiv
\frac{1}{2}\log {\cal F}.
\end{equation}
The unknown function $t_{\tilde{u}\tilde{u}}$ is to be determined by the
choice of the initial condition.
Alternatively, one may think of this flux as the S-wave sector contribution
to the actual four-dimensional flux, except that possible backscattering
effect is ignored.  Of course, one need to insert an appropriate scale
factor in order to recover four-dimensional S-wave flux density.

In any case, the only thing that remains to be done is choosing the initial
condition that will fix $t_{\tilde{u}\tilde{u}}$. Before we do this, however,
we first need to know what are the good coordinates near the horizons. As
in Ref.~\cite{none}, the approximate behavior of these Kruskal
coordinates near the horizons are completely determined by the respective
surface gravities. For the retarded time coordinates, specifically, we find,
\begin{equation}
U_{dS}\simeq \frac{1}{\kappa_{dS}}e^{\kappa_{dS} \tilde{u}},\qquad
U_{bh}\simeq -\frac{1}{\kappa_{bh}}e^{-\kappa_{bh} \tilde{u}}.
\end{equation}
Note that $U_{bh}$ vanishes at the future event horizon
($\tilde{u}=\infty$).

\vskip 1cm
\begin{center}
\leavevmode
\epsfysize=3.5in \epsfbox{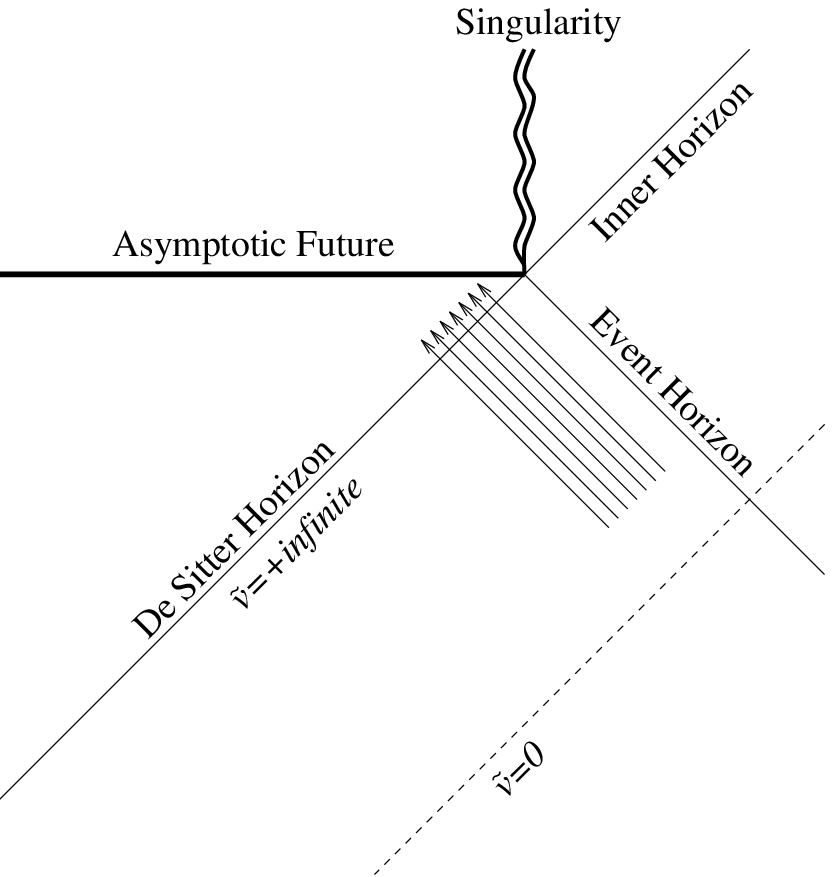}
\end{center}
\begin{quote}
{\bf Figure 4:} {\small Radiation from the black hole toward De Sitter
horizon ($\kappa_{dS} < \kappa_{bh}$) is depicted schematically.
The vacuum chosen is of the Unruh type.}
\end{quote}

\vskip 5mm

In order for the future event horizon to remain smooth at one-loop, it is
necessary that the local energy density (as measured by infalling inertial
observers) remains finite at one-loop. But the latter is given by the
quantity $\langle T_{\tilde{u}\tilde{u}}\rangle (d\tilde{u}/dU_{bh})^2$ near
the future event horizon, which means that this is possible only if $t_{
\tilde{u}\tilde{u}}$ cancels the leading contribution from the derivatives
of $\rho$, near the future event horizon $u\rightarrow \infty$. It turns out
that this prescription is also sufficient whenever $\kappa_{bh}> 0$.

Once we demand such an Unruh type initial condition on an initial surface,
say $\tilde{v}=0$, the late-time behavior of the outward flux $\langle T_{
\tilde{u}\tilde{u}} \rangle$ along $\tilde{v}=\infty$, is uniquely
determined:
\begin{equation}
\langle T_{\tilde{u}\tilde{u}}\rangle \sim \hbar\,
[\kappa_{bh}^2-\kappa_{dS}^2]
\quad \hbox{ as } \tilde{u}\rightarrow \infty. \label{toyH}
\end{equation}
If the De Sitter horizon were absent ($H=0$ so that  $\kappa_{dS}=0$),
the hypersurface $\tilde{v}=\infty$ would be the asymptotic future. In that
case, it is easy to see that Eq.~(\ref{toyH}) does reproduce the usual
late-time thermal radiation from the RN black hole, since the
radiative flux from a blackbody (in 1+1 dimension) of temperature $T$
scales like $T^2/\hbar$. Recall that the Hawking temperature is given by the
surface gravity $\kappa_{bh}$ multiplied by $\hbar/2\pi$.
The universal nature of the Hawking radiation is already apparent here, in
that the above behavior (\ref{toyH}) does not depend
on the details of the initial condition
other than the regularity of the future event horizon. Neglected here are
non-universal subleading terms that are exponentially small as $\tilde{u}
\rightarrow \infty$.

In the presence of the De Sitter horizon ($\kappa_{dS} >0$), however, there is
one more step to be taken. First of all, $\tilde{u}$ is not a good coordinate
at the future De Sitter horizon ($\tilde{v}=\infty$).
Using the local inertial coordinate $U_{dS}$ instead, the energy flux that
crosses the De Sitter horizon into the asymptotic future,
is given by the following,
\begin{equation}
\hbox{Outward Energy Flux}\simeq \langle T_{U_{dS}U_{dS}}\rangle\sim
\frac{\hbar \,[\kappa_{bh}^2-\kappa_{dS}^2]}{\kappa_{dS}^2U_{dS}^2}.
\end{equation}
as $U_{dS}\rightarrow\infty$. At last, we see
that the effect of the extra horizon on the Hawking radiation is two-fold.

The first  is the red-shift factor $(\kappa_{dS}U_{dS})^{-2} \simeq e^{-2
\kappa_{dS}\tilde{u}}$ that originates from the fact that the ambient
universe is undergoing an inflationary expansion. The physical distance
between the source (i.e., the black hole) and the inertial observer
propagating along a fixed advanced time outside the cosmological horizon,
grows exponentially with $\tilde{u}$, and so does the relative red-shift
factor. The effective temperature is red-shifted likewise by a factor
$(\kappa_{dS}U_{dS})^{-1}$. Is there an analogous effect in the case of
uniformly accelerated black holes?

In the Ernst spacetime, consider the inertial observers on the axis of
rotational symmetry and between the two accelerated RN black holes. If energy
quanta from the black holes reach them, the energy must be Doppler-shifted
due to the instantaneous relative motion. It is matter of a simple Lorentz
transformation to see that the appropriate red-shift factor
scales as $\sim e^{-\tilde{s}}$ at large dimensionless proper time
$\tilde{s} \simeq As$ of the black hole. Since the black hole is at the
Rindler coordinate $\eta\simeq 1/A$, we may rewrite this in terms of the
retarded time $(T-Z)\sim e^{-\tilde{s}}/A$ (see section 2 for the notation).
This tells us that the effective temperature would be red-shifted by the
factor $(A\,(T-Z))^{-1}$ to such inertial observers at the retarted time $T-Z$.
Identifying $A$ with $\kappa_{dS}$ and $T-Z$ with $U_{dS}$, this produces
the same kind of the red-shift factor as above.

The second, much less intuitive, effect is to replace $\kappa_{bh}^2$ by
$\kappa_{bh}^2- \kappa_{dS}^2$. This has little to do with the instantaneous
relative motions, as is easily seen from the fact that the same modification
exists in the static coordinate $(\tilde{u},\tilde{v})$. The physical
interpretation of this is that the presence of the extra horizon completely
altered the original thermal spectrum. In particular, when the two surface
gravities coincides ($\kappa_{bh}=\kappa_{dS}$), {\it the leading
Doppler-shifted Hawking radiation is completely turned off}. As was shown
in Ref.~\cite{none} and will be reexamined in the next section, the same
phenomenon occurs for certain accelerated black hole.

Also note that if $\kappa_{dS}$ is larger than $\kappa_{bh}$, this
asymptotic flux has a negative sign, meaning the black hole is actually
accreting energy rather than evaporating. There is nothing wrong with
this since the ambient space comes with a uniform density of energy in the
form of the cosmological constant. Furthermore, as we will see in Section 5,
the typical classical geometries are such that the Hawking temperature is
almost always larger than that of the extra horizon outside. Such accretion
processes, if any, are always stopped before the black hole deviates far
from the extremality.

\section{Vanishing Hawking Radiation}

Now back to the real model of the four-dimensional Ernst spacetime. In this
section, for the sake
of the completeness, we want to reiterate the main arguments of
Ref.~\cite{none} but in a slightly more detailed fashion.
Let us first define the surface gravities of the relevant horizons in Ernst
spacetime:
\begin{equation}
\kappa_{BH}\equiv \frac{F'(\tilde{r}_+)}{2},\qquad
\kappa_{A}\equiv -\frac{F'(r_{A})}{2},
\end{equation}
which are related to the temperatures by $T_{BH}=\hbar\kappa_{BH}/2\pi$ and
$T_A=\hbar\kappa_{A}/2\pi$. Note that when the size of the black holes are
relatively small ($r_\pm A \ll 1$), $A\simeq \kappa_A$ can be regarded as
the acceleration of the black hole.
Since we are mostly interested in the cases where the co-moving Rindler
observers find a complete thermal equilibrium, we will require that the
Hawking temperature be equal to the acceleration temperature. In terms of
the surface gravity, therefore, we demand that
\begin{equation}
\kappa\equiv \kappa_{BH}=\kappa_{A}.\label{kappa}
\end{equation}
In some cases, most notably when $r_+\gg r_-$ so that the black hole mass is
much larger than its charge, $\kappa_{BH} >\kappa_A$ is always true and this
constraint can never be met. When the non-extremal RN black holes in question
are sufficiently close to the extremality, on the other hand, it is possible
to achieve this fine-tuning \cite{piljin}. In fact, this constraint is
naturally imposed if the two black holes were pair-created via a
wormhole-type instanton \cite{garfinkle}.

\vskip 5mm

Having the Ernst metric instead of the toy model of the
previous section gives rise to several difficulties, mostly due to the
lack of the spherical symmetry and the complicated pattern of the Doppler
effects. Fortunately, it turned out that none of these problems matter if we
are interested in the null Hawking effect when $\kappa_{BH}=\kappa_A$,
provided that we work with the eigenmodes in the Rindler coordinate and
construct other inertial eigenmodes via the analyticity argument of
Unruh \cite{unruh}.

For this purpose, it is most convenient to introduce a tortoise-like
coordinate $z$ where
$\tilde{r}_+\le r \le r_A$,
\begin{equation}
z\equiv \int^r d\tilde{r}\,\frac{1}{F(\tilde{r})},
\end{equation}
which logarithmically approaches $-\infty$ at the event horizon and
$+\infty$
at the acceleration horizon. Consider the field equation of a free scalar
$\Psi$ that may have a quadratic curvature coupling.
\begin{equation}
\nabla^2 \Psi=M^2\Psi+\cdots .
\end{equation}
After rescaling the eigenmodes $\Psi^{(w,m)}$,
for each Rindler frequency $w>0$ and the quantized angular momentum $m$,
\begin{equation}
\Psi^{(w,m)}= e^{\mp iws}\,\frac{(1+rAx)}{r}\,[\,\Phi_{(w,m)}(r,x)
e^{im\phi}\,],
\end{equation}
we find the following equation that must be solved for the eigenmodes:
\begin{equation}
w^2\Phi_{(w,m)}+\frac{\partial^2}{\partial z^2}\Phi_{(w,m)}=F(r(z))\biggl\{
\frac{1}{r^2}\biggl[-\frac{\partial}{\partial x} G(x)
\frac{\partial}{\partial
x} +\frac{m^2\Lambda^4}{G(x)}\biggr]+U_{\rm eff} \biggr\}\,\Phi_{(w,m)}.
\label{eigen}
\end{equation}
$U_{\rm eff}$ is a bounded function of coordinates $z$ and $x$, and in
particular contains the salar mass term and the
possible curvature couplings. Note that
the right-hand-side of Eq.~(\ref{eigen}) has the overall factor $F(r(z))$
that vanishes exponentially $\sim e^{-2\kappa |z|}$ as $|z|\rightarrow
\infty$. In the same limit, $\Lambda$ is reduced to functions of $x$
only, $\Lambda \rightarrow \Lambda_A\equiv \Lambda(r=r_A)$ or $\Lambda
\rightarrow\Lambda_{BH}\equiv\Lambda(r=\tilde{r}_+)$.

For any mode with finite transverse physical momentum along $x$
and $\phi$ directions, the exponentially small factor $F(z)$ causes an
effective separation of variables occurs near the horizons. Then,
introducing two null coordinates $u=s-z$ and $v=s+z$ in {\large L} and
also in {\large R}, we find the following general behavior, near each
horizon, of the
future-directed Rindler eigen-modes $\Psi^{(w,m)}_L$ and $\Psi^{(w,m)}_R$
that have respective supports in either {\large L} or {\large R} only,
\begin{eqnarray}
\Psi_L^{(w,m)} \sim e^{-iwu}C_{\lambda m}(x) e^{im\phi}\hbox{ \ or }
e^{-iwv}C_{\lambda m}(x) e^{im\phi}\quad \hbox{ in L},& & \qquad
\Psi_L^{(w,m)} =0 \quad \hbox{ in R }, \label{modeL}\\
\Psi_R^{(w,m)} \sim e^{+iwu}C_{\lambda m}(x) e^{im\phi}\hbox{ \ or }
e^{+iwv}C_{\lambda m}(x) e^{im\phi}\quad \hbox{ in R},& & \qquad
\Psi_R^{(w,m)} =0 \quad \hbox{ in L }. \label{modeR}
\end{eqnarray}
The positive sign in (\ref{modeR}) is because $(u,v)$ grow toward past
rather than future in the region {\large R}.
The same set of symbols $C_{\lambda m}$ and $\lambda$ are used to denote
eigenfunctions and eigevalues for two different eigenvalue problems at
each horizon. The relevant operators are obtained from the one inside the
square bracket in Eq.~(\ref{eigen}), by replacing $\Lambda$ by $\Lambda_A$
or by $\Lambda_{BH}$. In particular, due to the lack of the spherically
symmetry, an eigenmode that has a definite $\lambda$ near the event
horizon will not have a definite $\lambda'$ near the acceleration horizon.
Similarly, possible backscattering will also mix the left-moving and the
right-moving modes. But since none of these details matter, as we will find
out shortly, we shall keep just one superscript $w$ form now on.

\vskip 5mm
Let us be reminded that, for each Rindler mode $\Psi^{(w)}$
with positive $w$, there exists a time-reversed negative mode $\Psi^{(-w)}$
that propagates backward but otherwise of the same form: The complete
Hilbert
space is spanned by both positive and negative modes. But the point is,
such labels as future-directed and past-directed are inherently
observer-dependent. A purely future-directed mode in one coordinate system
could be a mixture of both future-directed and past-directed modes as
perceived by another coordinate system. Hence, the so-called {\it Bogolubov
transformation}, which maps one basis to the other, is such that a vacuum
with respect to one set of observers can actually be an excited state with
respect
to the other. And this is exactly the origin of both the Hawking radiation
and the acceleration heat baths \cite{hawking}\cite{unruh}.
\vskip 1.5cm
\begin{center}
\leavevmode
\epsfysize 2.2in \epsfbox{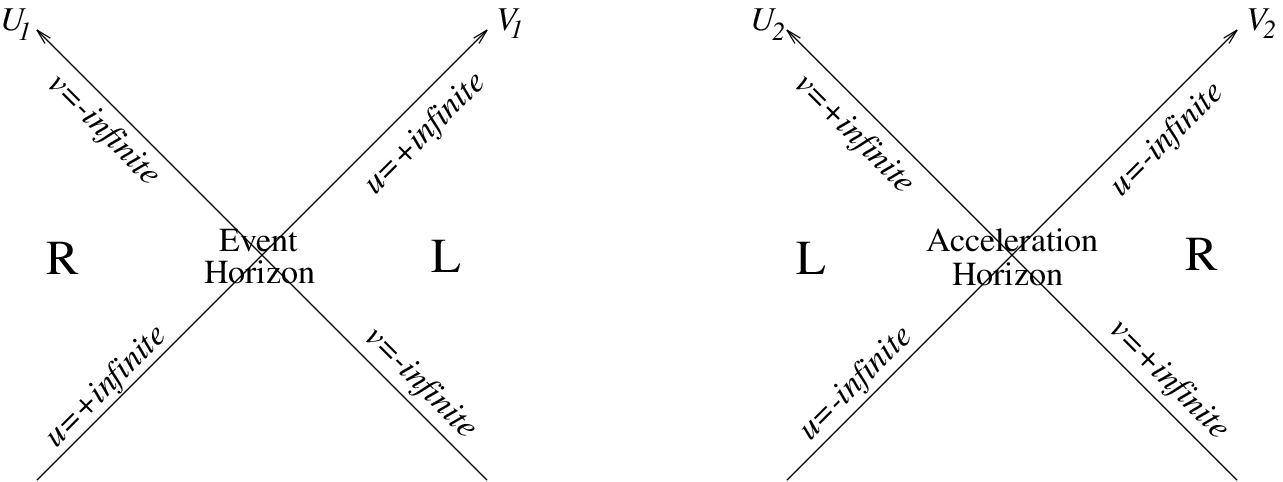}
\end{center}
\begin{quote}
{\bf Figure 5:} {\small Various null coordinates near the horizons. $U_1=0$
or $V_1=0$ at the event horizon, while $U_2=0$ or $V_2=0$ at the acceleration
horizon. All Kruskal coordinates increase toward future. The Rindler-type
null coordinates $(u,v)$, however, increase toward future only in L,  and
actually increase  toward past in R.}
\end{quote}
\vskip 5mm

In this regard, it is important to realize that $(u,v)$ are not good
coordinates near the horizons and must be traded off in favor of the Kruskal
coordinates that play the role of advanced and retarded times for local
inertial observers. The approximate coordinate transformation, near the
respective horizons, are easily determined in terms of the surface gravity
$\kappa$. Calling the Kruskal coordinates near the event horizon $(U_1,V_1)$,
we find,
\begin{eqnarray}
\kappa u\simeq - \ln (-U_1) \quad\hbox{ in L},& &\qquad
\kappa u\simeq - \ln (+U_1) \quad\hbox{ in R}, \label{uU}\\
\kappa v\simeq + \ln (+V_1) \quad\hbox{ in L},& &\qquad
\kappa v\simeq + \ln (-V_1) \quad\hbox{ in R}. \label{vV}
\end{eqnarray}
For the other Kruskal coordinates $(U_2,V_2)$ near the acceleration horizon,
we simply replace $(U_1,V_1)$ by $(U_2,V_2)$ and reverse every single sign
on the right-hand-side.

\vskip 5mm
At last, we are ready to obtain the eigenmodes of positive frequencies
with respect to inertial observers. For the purpose, we may use Unruh's
characterization of positive frequency \cite{unruh}: through a simple
analyticity argument, it is easy to see that a positive frequency mode must
be
analytic and bounded in the lower-half-plane of the complexified time
coordinate. For instance, a positive frequency mode as detected by inertial
observers near the event horizon must be analytic in $U_1$ (with fixed $V_1$)
and in $V_1$ (with fixed $U_1$) throughout their lower-half-planes. Fore
more detail, see Section II of Ref.~\cite{unruh}.

Since the Rindler modes are defined in either {\large L} or {\large R}, they
are defined only on the half-lines of Kruskal coordinates. To construct the
eigenmodes that are appropriate for inertial observers, one expresses
$\Psi^{(w)}_L$ and $\Psi^{(w)}_R$ in terms of $(U_i,V_i)$ for $i=1,2$ using
the coordinate transformations above in (\ref{uU}) and (\ref{vV}), and
analytically continue the logarithms through lower-half-planes of each
Kruskal coordinates. Then the resulting modes have positive frequencies with
respect to inertial observers, in addition to having the supports on the
entire spans of Kruskal coordinates. For inertial observers
near the event horizon of the black holes, the positive frequency modes
${\Psi}_B^{(w)}$ are
\begin{equation}
{\Psi}^{(w)}_{BL}\simeq N_w[\Psi^{(w)}_L+e^{-\pi w/\kappa} \Psi^{(-w)}_R],
\qquad
{\Psi}^{(w)}_{BR}\simeq N_w[\Psi^{(w)}_R+e^{-\pi w/\kappa} \Psi^{(-w)}_L],
\label{RB}
\end{equation}
where $N_w\equiv 1/\sqrt{1-e^{-2\pi w/\kappa}}$. Expressions for
the negative modes $\Psi^{(-w)}_B$ can be found likewise.

For an ordinary nonaccelerated black holes ($\kappa=\kappa_{BH}$,  $\kappa_A=
0$), this would be the end of the story: The null coordinates $(u,v)$ become
the asymptotic retarded and advanced time coordinates, so that requiring the
smooth future event horizon necessarily implies proliferation of particles
(associated with $\Psi^{(w)}_L$'s or $\Psi^{(w)}_R$'s here.) at
asymptotic infinities. The spectrum would follow the Bose-Einstein
distribution $\sim e^{-2\pi w/\kappa}N^2_w$ \cite{hawking}.

\vskip 5mm
However, with the uniformly accelerated black hole, $(u,v)$ are no longer
good asymptotic inertial coordinates. Rather, $(U_2,V_2)$ are. The geometry
(\ref{geometry}) becomes the background Melvin spacetime far away from the
black holes, but written in terms of Rindler-like coordinate. This is
particularly clear when the black holes are relatively small ($r_\pm A
\ll 1$), as we observed in section 2. The Minkowski-type coordinates
$T$ and $Z$ in (\ref{mink}), are related to our Kruskal coordinates
by $U_2=T-Z$ and $V_2=T+Z$, which makes it quite explicit that $U_2$ and $V_2$
are the retarded and the advanced time coordinates appropriate for
asymptotic inertial observers.

Near the acceleration horizon, the situation is identical to the above,
thanks to the identical surface gravity $\kappa_A=\kappa=\kappa_{BH}$,
except that
the relative positions of {\large L} and  {\large R} are switched. Calling
the asymptotic inertial modes $\Psi_A^{(w)} $'s, we find near the
acceleration horizon:
\begin{equation}
{\Psi}^{(w)}_{AR}\simeq N_w[\Psi^{(w)}_R+e^{-\pi w/\kappa} \Psi^{(-w)}_L],
\qquad
{\Psi}^{(w)}_{AL}\simeq N_w[\Psi^{(w)}_L+e^{-\pi w/\kappa} \Psi^{(-w)}_R].
\label{RA}
\end{equation}
When compared to (\ref{RB}), the particular thermal nature of this immediately
tells us that the Rindler observers must find some sort of thermal equilibrium.
On the other hand, the Bogolubov transformation relevant for the asymptotic
observers is found by combining (\ref{RB}) and (\ref{RA}), but this does not
lead to any mixing between the positive and negative modes:
\begin{equation}
\Psi^{(w)}_B\Rightarrow \Psi^{(w)}_A,\qquad
\Psi^{(-w)}_B\Rightarrow \Psi^{(-w)}_A. \label{main}
\end{equation}
Therefore, we find that {\it no late-time Hawking radiation reaches the
asymptotic inertial observers}.

\vskip 5mm
As noted earlier, the respective forms of the Rindler modes near each
horizon must be taken with a grain of salt, since the nontrivial effects
of the local geometry mixes up left-moving modes with the right-moving modes
and also require rather complicated $x$-dependence near at least one of the
horizons. Combining the two Bogolubov transformations, in general, we must
include a unitary transformation $\cal U$ that reflects the complicated effects
of local geometry between the two horizons. But, since the Rindler time $s$
is a Killing coordinate, $\cal U$ has to commute with the Rindler energy
operator $i\partial_s$. Therefore, with the Bogolubov transformations above
that depend only on the Rindler frequency $w$, this additional complication
cannot alter our conclusion above.

\vskip 5mm
Actually there is a more direct way to understand this result. Consider
the direct transformations between the two sets of Kruskal coordinates
when the two surface gravities match:
\begin{equation}
U_2\sim -\frac{1}{U_1}, \qquad V_2\sim -\frac{1}{V_1}.\label{ex}
\end{equation}
This transformation is easily seen to
preserve the lower-half-planes of the Kruskal time
coordinates, being an $SL(2,R)$ generator, which in turn explains why
the Bogolubov transformation thereof is trivial to the leading
approximation. In fact, this transformation (\ref{ex}) is exactly what we
would have found if we had been considering a freely falling extremal RN
black hole that has zero Hawking temperature and thus no late-time Hawking
radiation, provided that we again identify $U_2$ and $V_2$ as the asymptotic
inertial time coordinates.

This last observation provides us with yet another  strong evidence that
our conclusion is correct for a single accelerated black hole as well.
For real extremal black holes, the above argument that utilizes the
property of $SL(2,R)$ transformations is not entirely correct since the
relevant range of $U_1$ for instance should be confined to the
negative real line. Nevertheless, the
fact that the extremal black hole is of zero Hawking temperature remains true:
While the resulting Bogolubov transformation is not entirely trivial, it
does not involve a steady flow of radiation energy either. In a similar
vein, we expect the present coordinate transformation in (\ref{ex})
lead to the vanishing Hawking radiation for the single accelerated black hole
although only half-lines of the Kruskal coordinates overlap with each other.

As this discussion shows, there is some subtlety that remains to be
addressed. The above derivation is fine as long as the late-time
thermal behavior of the Hawking radiation is concerned, but there could be
in general certain transient radiation which has not been precluded by this.
Exactly how does such a nontrivial result show up beyond the trivial late-time
 Bogolubov transformation above? The effect comes from at least
two different sources. The obvious one is the fact that the above coordinate
transformation is only approximate and is valid very close to the horizons.
The less obvious factor, which one can easily overlook, is that the ranges of
different Kruskal coordinates above overlap with each other completely
only in the maximally extended spacetime. For more realistic cases without
past event horizon, it is typically the case that only the half-lines of
each Kruskal coordinates are relevant, as briefly mentioned in the previous
paragraph, and this leads to certain nontrivial result.

Fortunately, none of these complicated effects introduce an analogue
of the Hawking radiations.
Rather, they merely lead to a finite and small one-loop correction of the
black hole geometry. In the toy model of Section 3, this kind of transient
behavior could be found in the subleading contributions  which are suppressed
by at least two more
powers of $\kappa_{bh}U_{bh}\simeq -e^{-\kappa_{bh}\tilde{u}}
$ as $\tilde{u}\rightarrow \infty$, and which depends on the detailed history.
In fact, such an effect that shifts the black hole mass by a finite and
small factor $\sim \hbar/r_+^2$, was previously observed for static extremal
RN black hole \cite{jaemo} directly as well as for the case of pair-produced
near-extremal RN black holes \cite{piljin} indirectly. Once a steady
state (with $\kappa_A=\kappa_{BH}$) is reached, in any case,
the main result (\ref{main}) tells us that no further radiation may escape
into the asymptotic future beyond the acceleration horizon.

\section{Evolution}

Let us consider the Ernst spacetime (\ref{geometry}) at arbitrary
accelerations. The first fact we need to understand, is that the acceleration
cannot really be arbitrary. The interpretation of this geometry as
a pair of black holes presumes that there are at least three distinct
roots of the quartic polynomial $G$, which is  false for some values
of $r_+A$.

The necessary and sufficient condition for the accelerated black
hole interpretation, turned out to be $r_+A< 2/\sqrt{27}$ \cite{gauntlett}.
The physical reason is clear: As either the black hole size or the strength
of the acceleration becomes too large, the event horizon and the acceleration
horizon eventually have to merge with each other and then subsequently
disappear. When they merge ($r_+A=2/\sqrt {27}$), we find $\xi_2=\xi_3=-
\sqrt 3$ and $\xi_4=\sqrt 3/2$. For the rest of the section, we shall work
with the harmless assumption that the smallest root $\xi_1$ is given by
$-1/r_-A$. Even if this is not the case, it does not alter our conclusion
as long as $r_-A$ is a positive number. The latter is again required for
the physical interpretation of the Ernst metric.

In general Ernst geometry, the ratio between the two temperatures can be
conveniently written as follows,
\begin{equation}
\frac{T_A}{T_{BH}}=\frac{\kappa_A}{\kappa_{BH}}=\frac{(\xi_4-\xi_3)
(\xi_3-\xi_1)}{(\xi_4-\xi_2)(\xi_2-\xi_1)}.\label{ratio}
\end{equation}
As we decrease $r_+A$ from $2/\sqrt{27}$, $\xi_4$ spans between $\sqrt 3/2$
and $1$, and $\xi_3$ between $-\sqrt 3$ and $-1$, which means that $(\xi_4-
\xi_3)$ is always of order one. Now, there are two different cases when this
ratio is one: either $\xi_2=\xi_3$ or $(\xi_2-\xi_1)=(\xi_4-\xi_3)\sim
1$. The first is the unphysical case where the two horizons merge while
the second, for $r_-A < \tilde{r}_+ A \ll 1$,
can be translated into the following,
\begin{equation}
\frac{({r}_+ -r_-)}{4\pi {r}_+^2}\simeq \frac{A}{2\pi},
\end{equation}
which is clearly the condition of equal temperatures.

When can $T_A$ be larger than $T_{BH}$? One must have either $(\xi_4-\xi_3)
>(\xi_4-\xi_2)$ or $(\xi_4-\xi_3) > (\xi_2-\xi_1)$, but the first cannot be
satisfied simply bacause $\xi_2 < \xi_3$ by definition. On the other hand,
the latter condition can be met only for those black holes sufficiently near
the extremality. Why? In order to be far from the extremality, $Ar_-=-1/\xi_1$
must be much smaller than $A\tilde{r}_+=-1/\xi_2$. But this implies $(\xi_2-
\xi_1)\gg 1 \sim (\xi_4-\xi_3)$, which together with $\xi_2 < \xi_3$ tells
us that the ratio $T_A/T_{BH}$ in Eq.~(\ref{ratio}) is necessarily smaller
than the unity.
In particular, for relatively small black holes ($r_-A<\tilde{r}_+ A \ll 1$),
it is easy to see that $T_A/T_{BH}$ monotonically decreases as a function
of the black hole mass and equals to one when $({r}_+-r_-)/r_+\simeq 2r_+A
\ll 1$. Note that the heat capacity of the black hole is positive if
$({r}_+-r_-)/r_+ \ll 1$.

In short,  {\it black holes far from the extremality
will always have its Hawking temperature larger that the
acceleration temperature}. Actually, this property appears quite
general and not very specific to the particular solution we are using.
For instance, if we replace the Ernst metric by the C-metric \cite{C} where the
black holes are accelerated by a pair of string-like singularities instead,
we find the exactly same phenomenon.\footnote{The C-metric is obtained by
setting $\Lambda=1$ in the metric (\ref{geometry}) and readjusting the
periodicity of $\phi$ coordinate.} Therefore, black holes that are sufficiently
far away from the extremality must continue to evaporate (with a somewhat
modified spectrum) even during the acceleration. In particular, this suggests
that the Schwarzschild black hole cannot be stabilized by a uniform
acceleration.

This observation immediately precludes a possible disaster. Suppose
that the black hole actually absorbs net energy from the acceleration
heat bath when $T_{BH}<T_A$. What happens if any RN black hole (of negative
heat capacity) can undergo an arbitrarily large acceleration? The black hole
would begin to accrete energy quanta from the acceleration heat bath.
With a fixed external force that drives the acceleration, $T_A$ may decrease
due to the increasing mass, but so should $T_{BH}$ at a similar rate
due to the negative heat capacity. Then, the process might continue
forever, leading to an arbitrarily massive black hole.

In contrast, near-extremal black holes with positive heat capacity may
undergo a large acceleration ($T_A>T_{BH}$); but the subsequent
accretion of mass must increase $T_{BH}$ and at the same time
decrease $T_A$. After a while,  the evolution will take the state to
a thermal equilibrium at a new temperature that lies in between the two
original temperatures.

The upshot is that {\it the uniform acceleration does not introduce any
runaway behavior} that involves a non-extremal
RN black holes accreting unlimited
amount of energy quanta from the acceleration heat bath. Instead, it simply
shifts the ground state from the extremal black hole to a near-extremal
variety whose Hawking temperature is equal to the acceleration temperature.
Similar considerations may be applied to the toy model of section 3 with
the same conclusion.

\section{Discussions}

One might ask whether the eternal uniform acceleration of the black hole
is really physical. Should we not consider only finite processes? In fact,
such skepticisms surfaced time and again in regard to the superficially similar
classical system of a uniformly accelerated charge. As early as 1910's,
physicists debated whether a uniformly accelerated charge really emits
Bremmstrahlung or not. The problem was that the uniformly acceleration does
not involve any radiation backreaction on the charge itself, which seemed to
suggest that there is no source for the radiation energy. In contrast,
as soon as one considers only finite processes where the charge experiences
a finite net momentum transfer, the integrated radiation energy matches
exactly the total work done by the radiation backreaction. Maybe it is not
``legitimate'' to consider such infinite processes within the framework
set by the classical electrodynamics, one may argue.

However, the classical physics of charged particle undergoing an eternal
uniform acceleration has been thoroughly understood by Boulware in early 80's
\cite{boulware}: The uniformly accelerated charge does emit Bremmstrahlung
into the asymptotic future across the acceleration
horizon, but the radiation energy does not originate from the
charge. Rather the destructive interference between the Coulomb field and the
radiation field reduces the field energy near the charge just the right amount
so that the net radiation energy is entirely explainable without violating
energy conservation. Whenever the uniform acceleration is disrupted, the
radiation backreaction transfer some energy from the charge to the surrounding
Coulomb field again, but during the uniform acceleration the total energy of
the electromagnetic field remains constant \cite{coleman}.\footnote{With
appropriate regularization of the point-like charged particle, of course.}

\vskip 5mm
Throughout this paper, we have been asking what happens quantum mechanically
if we replace the charge by a uniformly accelerated charged black hole.
Obviously, the classical radiations associated with electromagnetic (and
gravitational) fields must be still observable to freely falling observers,
since classical observers at large spatial distances cannot distinguish
a point-like source from a black hole. But what about the one-loop effects?

There is one particular case where the analogy with the above classical
system appears almost complete, that is, when the co-moving
observers find an equilibrium state even at one-loop level, on account of the
fact that the Hawking temperature and the acceleration temperature are
fine-tuned to be equal. If we had extended the analogy all the way, we would
have believed that the inertial observers detects the black hole radiance
at arbitrary late-time while co-moving observers may still insist upon no
evaporation. However, a careful one-loop analysis above revealed that this is
not the case at all. All observers actually agree when it comes to the Hawking
radiation: There is none whatsoever.

It would have been a very difficult situation, if different observers
found different evolutions of the black hole. It would have meant among
others that the mass of the black hole is observer-dependent and maybe
so is the Bekenstein-Hawking entropy. While such  behaviors are not
entirely inconceivable, considering the complicated nature of the gravity,
still it is  gratifying to know that the simplest explanation available is
also the correct one.

\vskip 5mm
One of the more important implications of our findings above, concerns the
natural vacuum state associated with the Ernst geometry. Since the Euclidean
version of the geometry, when $\kappa_{BH}=\kappa_A$, is the wormhole-type
instanton that pair-creates near-extremal RN black holes \cite{garfinkle},
the matter of the natural vacuum is of fundamental importance in the WKB
estimate of the tunneling rates \cite{gauntlett}\cite{piljin}.

Recall that the natural vacuum of a ``freely falling'' Euclidean black hole
is the so-called Hartle-Hawking state, which entails an asymptotic heat bath
of $T_{BH}$. If one insists upon a zero-temperature state at large distances
(in order to maintain finite total energy, for instance), such as the
Boulware state, this leads to a severe divergence of the energy-momentum
expectation at the event horizon. And this happens precisely because of the
nontrivial Bogolubov transformation that relates the asymptotic inertial
observers to their counterpart near the event horizon.

In contrast, our findings above imply that a vacuum state of
the Ernst space may resemble the Hartle-Hawking vacuum near the black hole
yet asymptotically behaves as an zero-temperature state of Boulware type.
It is as if the acceleration heat baths that surround the pair-created
black holes are of a naturally finite size. In a previous work \cite{piljin}
on the matter of RN black hole pair-creation, the author has asserted that
the natural vacuum is characterized precisely by such properties.

Why is this vacuum state especially important for the semiclassical estimate
of the tunneling rate? Note that, within the semiclassical framework where
the classical geometry is treated also as a dynamical object,
a consistent WKB expansion must include the gravitational backreaction to
the quantum energy-momentum. That is, in order to obtain a consistent
$(n$+$1)$-th order WKB estimate of the tunneling rate, we must first obtain
the quantum-corrected instanton geometry that solves the Euclidean Einstein
equation with the total energy-momentum to $n$-th order. Naively, up to the
one-loop order, one may start with the zero-th order classical Euclidean
instanton solution.

Now suppose that the natural Euclidean vacuum state involved an asymptotic
heat bath, as one would expect for a stationary black hole. Although the
one-loop energy density could be made arbitrarily small by considering large
black holes with small Hawking temperatures, the total energy thereof would
always diverge since the heat bath should entail a finite uniform energy
density in the asymptotic region. Then, the one-loop corrected instanton would
have a very different asymptotic geometry from the one we started with, and it
becomes rather unclear whether and in what sense the quantum correction might
be small enough to justify the WKB expansion to begin with. More specifically,
while it is conceivable that there exists a physical prescription of
performing a systematic expansion in this case,
it is hardly obvious whether one will obtain
the right answer from the naive approach based on the classical instanton
solution, e.g., as in Ref.~\cite{why}.

Fortunately, with the present modified vacuum, the heat bath is
cut off essentially where the Euclidean black hole is truncated, that is,
at $r^3 \sim r_+/B^2$ in the notation of section 2. The total one-loop energy
must be finite, and in the weak field limit or equivalently in the small
acceleration limit, the dominant contribution is near the Euclidean black
hole. Subsequently, the only gravitational backreaction to such one-loop
energy-momentum is to change the local geometry near the black  hole a
little bit, and in particular manifest itself in a small one-loop
correction to the black hole mass of order $\sim \hbar/r_+$ while the
temperature is held fixed \cite{piljin}. For small enough $\hbar/r_+^2$,
therefore, we find that the systematic WKB approximation based on the
classical instanton is sensible thanks to this unusual nature of the vacuum.

\vskip 5mm
In summary, we studied the semiclassical evolution of accelerated
Reissner-Nordstrom black hole, and found that the uniform acceleration
actually shift the final ground state from the extremal to a near-extremal
variety. A similar phenomenon is also found in the toy model of charged
black holes in a dimensionally reduced De Sitter universe.
Our findings also provide an important consistency check for
the WKB expansion of the RN black hole pair-creation tunneling rates.

\end{document}